\documentclass[prd,onecolumn,amsmath,amsfonts,amssymb]{revtex4}
\usepackage{epsfig,psfrag,graphicx,bm,color,verbatim}

\begin{document}

\rightline{FTUAM-11-56}
\rightline{IFT-UAM/CSIC-11-70}
\rightline{October 2011}

\vspace{0.7cm}

\title{CLUES on Fermi-LAT prospects for the extragalactic detection of $\mu\nu$SSM gravitino Dark Matter}

\author{G.~A.~G\'{o}mez-Vargas$^{1,2,3}$, M.~Fornasa\footnote{MultiDark fellow}$^{4}$, F.~Zandanel$^{4}$, 
A.~J.~Cuesta$^{5}$, C.~Mu\~{n}oz$^{1,2}$, F.~Prada$^{4}$ and G.~Yepes$^{1}$}

\affiliation{${^1}$ Departamento de F\'{\i}sica Te\'{o}rica, Universidad Aut\'{o}noma de Madrid, Cantoblanco, E-28049, Madrid, Spain}
\affiliation{${^2}$ Instituto de F\'{\i}sica Te\'{o}rica IFT-UAM/CSIC, Universidad Aut\'{o}noma de Madrid, Cantoblanco, E-28049, Madrid, Spain}
\affiliation{${^3}$  Istituto Nazionale di Fisica Nucleare, Sez. Roma Tor Vergata, Roma, Italy}
\affiliation{${^4}$ Instituto de Astrof\'{\i}sica de Andaluc\'{\i}a (CSIC), E-18008, Granada, Spain,} 
\affiliation{${^5}$ Department of Physics, Yale University, New Haven, CT 06511, USA} 


\begin{abstract}
The $\mu\nu$SSM is a supersymmetric model that has been proposed
to solve the problems generated by other supersymmetric extensions of the standard model of particle physics. Given that R-parity is broken
in the $\mu\nu$SSM, the gravitino is a natural candidate for decaying dark matter since its lifetime becomes much longer than the age of the Universe. In this model, gravitino dark matter could be detectable 
through the emission of a
monochromatic gamma ray in a two-body decay.
We study the prospects of the Fermi-LAT telescope to detect such 
monochromatic lines in 5 years of observations of the most massive nearby
extragalactic objects. The dark matter halo 
around the Virgo galaxy cluster is selected as a reference case, since it is associated to
a particularly high signal-to-noise ratio and is located in a region
scarcely affected by the astrophysical diffuse emission from the galactic plane.
The simulation of both signal and background gamma-ray events is carried out 
with the Fermi Science Tools, and the dark matter distribution around Virgo is 
taken from a $N$-body simulation of the nearby extragalactic Universe, with constrained initial conditions provided 
by the CLUES project.
We find that a gravitino with a mass range of $0.6$--$2$ GeV, and with a lifetime range of about 
$3\times 10^{27}$--$2\times10^{28}$ s would be
detectable by the Fermi-LAT with a signal-to-noise ratio larger than 3.
We also obtain that gravitino masses larger than about 4 GeV are already excluded in the $\mu\nu$SSM by Fermi-LAT data of the galactic halo.

\end{abstract}
 
\maketitle
\section{Introduction}
Evidences indicating the presence of dark matter (DM) can be obtained at 
very different scales: from cosmological ones through the analysis of the 
angular anisotropies in the cosmic microwave background radiation \cite{Komatsu:2010fb}, down to 
galactic scales considering lensing and galaxy dynamics studies. 
However, all these evidences are not able to provide us with complete 
information on the nature of DM, beyond the fact that it has to be 
mainly non-baryonic.
Since,
within the standard model of particle physics there are 
no viable non-baryonic candidates, 
the existence of DM represents one of the most compelling evidences for 
physics beyond the standard model \cite{reviews}.

The ``$\mu$ from $\nu$'' Supersymmetric Standard Model ($\mu\nu$SSM) 
was proposed in the literature to solve the so called $\mu$-problem 
and explain the origin of neutrino masses by simply introducing right-handed 
neutrinos $\nu$ \cite{LopezFogliani:2005yw,reviewsmunu}.
It is an interesting model that can be tested at the LHC and, as a 
consequence, its phenomenology has been analized in detail recently 
\cite{Escudero:2008jg,Choi:2009ng}. 
In the $\mu\nu$SSM, $R$-parity is broken and therefore the lightest 
supersymmetric particle (LSP) decays. Thus, neutralinos \cite{reviews} 
or sneutrinos \cite{sneutrino}, having very short lifetimes, are no longer 
viable candidates for the DM of the Universe. Nevertheless, if the role
of the LSP is played by the gravitino, $\Psi_{3/2}$, its decay is suppressed 
both by the feebleness of the gravitational interaction and by the small 
$R$-parity violating coupling.
As a consequence, its lifetime can be much longer than the age of Universe and 
the $\mu\nu$SSM gravitino can represent a good DM candidate \cite{Choi:2009ng}.

Since the gravitino decays producing a monochromatic photon with an energy 
equal to half of the gravitino mass, its presence can, in principle, be 
inferred indirectly from the data of
the Fermi gamma-ray space telescope \cite{Choi:2009ng}.
Fermi was launched on June 11th, 2008, and its main instrument, the Large
Area Telescope (LAT), covers an energy range from roughly 20 MeV to 300 GeV 
with an angular resolution of $\sim$ 0.15 degrees at $68\%$ containment above 10 GeV 
\cite{Atwood:2009ez}.

In the DM halo of the Milky Way the gamma-ray flux coming from DM decay is 
maximized in the direction of the galactic center where the DM density is 
larger. This region, however, should be considered with particular care 
since the gamma-ray emission due to the presence of conventional astrophysical sources 
is not fully understood \cite{Lapi:2009ee}. 

The expected diffuse gamma-ray emission from DM decay in the 
mid-latitude range ($10^\circ \leq |b| \leq 20^\circ$) was computed for a Navarro-Frenk-White (NFW) profile \cite{navarro} in 
Ref. \cite{Choi:2009ng} and compared with the 5-month measurement reported 
by Fermi-LAT \cite{Abdo:2009mr}. The non-observation 
of sharp monochromatic lines in the gamma-ray spectrum permitted to draw 
bounds on the parameter space of the $\mu\nu$SSM gravitino.
In particular, 
values of gravitino mass $m_{3/2}$ larger than about 10 GeV were excluded, as 
well as lifetimes $\tau_{3/2}$ smaller than about 3 to 5$\times 10^{27}$ s.
Notice that because of this upper bound on $m_{3/2}$, three body decay 
modes of the gravitino \cite{Choi:2010jt,aurelio} are not relevant, and therefore we do not considered them in this work.

It is worth noticing here that in Ref. \cite{Abdo:2010nc} (\cite{Bloom})
the Fermi-LAT collaboration presented constraints on monochromatic emission 
using 11 (23) months of data for $|b| > 10^\circ$ {\it plus} a $20^\circ \times 20^\circ$ square around the galactic center. However, the derived limits only refer to 
the emission above 30 (7) GeV, covering, in the context of the $\mu\nu$SSM, 
gravitinos with masses larger than 60 (14) GeV, leaving our region of 
interest unconstrained. 
On the other hand, in work \cite{Vertongen:2011mu}, two-years Fermi-LAT data for $|b| \geq 10^\circ$ have been
used to constrain the DM gamma-ray line flux in the energy range between 1 
and 300 GeV. Lower bounds on $\tau_{3/2}$ of about $5 \times 10^{28}$ s 
were obtained in our region of interest below 10 GeV.
Recently, these bounds together with those obtained in \cite{kamion} by analyzing the data from EGRET were used in \cite{bilinear} to constrain the parameter space of gravitino dark matter in the
bilinear $R$-parity violating model.

We know from X-ray and lensing studies \cite{Markevitch:2001ri,Clowe:2006eq} that the main mass 
component of structures like galaxy clusters is DM. Such objects are, therefore, 
promising targets for DM searches. The main goal of this paper is to 
explore the prospects for detecting $\mu\nu$SSM gravitino DM in galaxy 
clusters using the prediction for 5 years of operation of the Fermi-LAT 
telescope, and taking into account the above mentioned bounds.
Although in a very recent work \cite{veryrecent} based on three years of Fermi-LAT gamma-ray data, the flux coming from nearby clusters was analyzed, the relevant bounds obtained on the lifetime correspond to gravitino masses larger than 250 GeV.

In our analysis we will use the following strategy. The DM density field of 
the nearby extragalactic Universe is described using the maps provided
in Ref. \cite{Cuesta:2010ex}, and based on a constrained $N$-body simulation 
provided by the CLUES project \cite{CLUES}.
This density is then taken as an input for the Fermi observation simulation 
tool to predict the photon signal. We will use the most recent version of the 
public Fermi Science Tools \cite{Tools} to describe the performance of the 
telescope and to simulate both the DM signal and the astrophysical background.
From the analysis of the simulated photon maps we finally compute our
prospects of detection for the $\mu\nu$SSM gravitino.

The paper is organized as follows. In Section~\ref{sec:two}, the flux of 
gamma rays from $\mu\nu$SSM gravitino decay is discussed. In 
Section \ref{sec:three}, the cosmological simulation used to infer the DM 
distribution in the Local Universe is briefly described, while 
Section~\ref{sec:four} is devoted to summarize the basic of the simulation 
technique implemented in the Fermi Science Tools. Finally, the prospects 
for DM detection are derived in Section~\ref{sec:five}, and the conclusions 
are left for Section~\ref{sec:six}.

\section{Gamma-rays from gravitino decay in the $\mu\nu$SSM}
\label{sec:two}

In the supergravity Lagrangian an interaction term is predicted between
the gravitino, the field strength for the photon, and the photino. 
Since, due to the breaking of R-parity, the photino and the left-handed 
neutrinos are mixed, the gravitino will be able to decay through the
interaction term into a photon and a neutrino \cite{Takayama:2000uz}.
The gravitino lifetime $\tau_{3/2}$ results to be:
\begin{equation}
\label{master}
\tau_{3/2} \simeq 3.8\times 10^{27} \mbox{s}
\left( \frac{\vert U_{\tilde{\gamma}\nu} \vert^2}{10^{-16}} \right)^{-1}
\left( \frac{m_{3/2}}{10 \mbox{ GeV}} \right)^{-3},
\end{equation}
where $\vert U_{\tilde{\gamma}\nu} \vert^2$ is the photino content of the neutrino, 
and is constrained to be $|U_{\widetilde{\gamma}\nu}|^{2}\sim 10^{-16}-10^{-12}$ in the $\mu\nu$SSM, in 
order to reproduce neutrino masses \cite{Choi:2009ng}. As a 
consequence, the gravitino will be very long lived.
Additionally, adjusting the reheating temperature one can
reproduce the correct relic density for each possible value of the gravitino 
mass (see \cite{Choi:2009ng} and references therein).

The detection of DM in several R-parity breaking scenarios has been studied 
in the literature \cite{Takayama:2000uz,Buchmuller:2007ui,Choi:2009ng} 
considering the case of gravitinos emitting gamma-rays when decaying in $i)$ 
the smooth galactic halo, and $ii)$ extragalactic regions at cosmological 
distances.
As mentioned in the Introduction, in this work we will analyze the detection 
of DM in the $\mu\nu$SSM considering the contribution of gamma-rays coming 
from $iii)$ nearby extragalactic structures.


In $i)$, the gamma-ray signal is an anisotropic sharp line and the flux is 
given by
\begin{equation}
\label{halo}
\frac{d\Phi}{dE}(E)=\frac{\delta(E-\frac{m_{3/2}}{2})}{4 \pi \tau_{3/2} m_{3/2}}
\int_{\textrm{los}}\rho_{halo}(\vec{l})d\vec{l}\ ,
\end{equation}
where the halo DM density is integrated along the line of sight $l$, and we 
will use a NFW density profile for the Milky Way halo compatible with the latest observational constraints as modeled in \cite{Prada:2004pi}. 
Let us remark, nevertheless, that since the region that we will study below does not include the inner galaxy, any density profile will give rise to similar results. 

On the other hand, in $ii)$, the photons produced by gravitinos decaying at 
cosmological distances are red-shifted during their journey to the observer, 
and we obtain the isotropic extragalactic flux applying the analysis
of Refs.~\cite{Takayama:2000uz,Overduin:2004sz} to the $\mu\nu$SSM.
As can be seen e.g. in Figs. 3 and 4 of Ref. \cite{Choi:2009ng}, the sharp 
line produced by the galactic halo dominates over this extragalactic signal.

Finally, for nearby extragalactic structures $iii)$, the gamma-ray signal 
is a monochromatic line similarly to $i)$, and Eq.~(\ref{halo}) can also be 
used for the computation of the flux. 
Actually the contribution from the smooth galactic DM halo is practically isotropic in the region around a particular nearby extragalactic structure (at least at high latitudes) and less important than the contribution from the object itself. Moreover at high latitudes, the galactic foreground is smaller than near the galactic center, so that objects with a lower DM-induced gamma-ray flux can potentially be associated to larger prospects of detection than the region near the galactic center. Thus, the study of the extragalactic density field is something worth carrying out.
In our work this is taken from the results of Ref. \cite{Cuesta:2010ex}, 
where an $N$-body simulation with constrained initial condition was used.
We will review the most important features of this simulation in the next 
Section.


\section{CLUES simulations}
\label{sec:three}
CLUES (Constrained Local UniversE Simulations) \cite{CLUES} $N$-body 
simulations aim at describing the formation and evolution of DM halos in a 
way to reproduce, as precise as possible, our Local Universe. To this goal, 
constrained initial conditions are set up using the information from radial
and peculiar velocities of galaxies from astrophysical catalogs, together 
with the determination of the masses of the galaxy clusters detected in 
X-rays \cite{Gottloeber:2010gv, Klypin}.

In Ref. \cite{Cuesta:2010ex}, a CLUES simulation was used to compute 
projected Cartesian grid all-sky maps of the DM distribution in the Local 
Universe. In particular, the Box160CR simulation was used, containing
1024$^3$ particles in a box with a side of 160 $h^{-1}$ 
\cite{Cuesta:2010ex,Gottloeber:2010gv}. 
The characteristics of the most massive clusters such as Virgo, Coma and 
Perseus, together with the Great Attractor, are well reproduced compared 
to the real objects, apart from a typical mismatch around 5 Mpc $h^{-1}$ in 
their position. 
The all-sky maps are available at Ref. \cite{Cuesta_maps} both in the case of 
an annihilating and a decaying DM particle. Here we use the map corresponding 
to the case of decaying DM. The gamma-ray flux can be derived from the 
values in the map simply multiplying them by the particle physics factor 
shown outside the integral in Eq. (\ref{halo}).
This result is used as input for the Fermi-LAT observation simulations, which 
we will describe in the next section.


\section{Simulations with the Fermi Science Tools}
\label{sec:four}

For this work, the simulation of gamma-ray events was carried out with 
the \texttt{gtobssim} routine, part of the Fermi Science Tools package 
v9r23p1. Its output is a list of mock gamma-ray events with corresponding spatial 
direction, arrival time and energy, distributed according to an input source 
model.
Our source model accounts for the gamma-ray signal from $\mu\nu$SSM 
gravitino decay as described in previous sections, and for the galactic and 
extragalactic background diffuse emission.
In particular, the galactic background emission is mainly correlated with 
structures in the Milky Way since it arises from the interaction of 
high-energy cosmic rays with the interstellar medium and the interstellar 
radiation field. The far extragalactic background, on the contrary, is 
supposed to be almost isotropic. Its value is based on the modelization of 
the galactic component, on detected Fermi-LAT sources, and on the solar 
gamma-ray emission. 
We used the so-called RING model \cite{Background} as recommended by the 
Fermi-LAT collaboration, which is obtained as a fit to the real Fermi-LAT
data.

The simulation of gamma-ray events through \texttt{gtobssim} is based on
in-flight Instrument Response Functions (IRFs), accounting for the telescope 
effective area, energy dispersion and point-spread function (PSF).
Two IRFs publicly available are called P6\_V3\_DIFFUSE
and P6\_V3\_DATACLEAN. They both profit from the improvement in the knowledge 
of the telescope performances after the first two years of data taking
\cite{Rando:2009yq}.
The main difference between the two IRFs is the fact that DATACLEAN event 
selection perform most stringent cuts than DIFFUSE on the interpretation of
an event as a real photon.
As will be discussed below, in this work we present results from P6\_V3\_DIFFUSE, but the sensitivity of 
our results on the choice of the IRFs have been checked without finding any significant effect.

\section{$\mu\nu$SSM gravitino dark matter: prospects of detection with Fermi-LAT}
\label{sec:five}

\begin{figure}[t]
\includegraphics[width=0.9\textwidth]{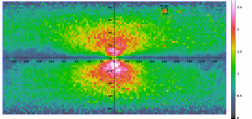}
\caption{\label{fig:one} $S/N$ all-sky map of gamma-ray emission from gravitino DM decay simulated using the Fermi Science Tools. 
DM events are from the decay of a $\mu\nu$SSM gravitino with $m_{3/2}=8$ GeV and $\tau_{3/2}=2.5 \times 10^{26}$ s. The signal comes from all three DM components: galactic halo, and both, local (from CLUES simulation) and far (isotropic) extragalactic DM density. The black square 
indicates the position of the Virgo cluster. Pixels in the map have an angular dimension of $3\times 3$ degrees.}
\end{figure}

Let us finally study the prospects for $\mu\nu$SSM gravitino DM detection, 
taking into account the contributions discussed in Sect.~\ref{sec:two}.
First, we simulate the all-sky map of 5-years Fermi-LAT observations, 
separately for DM gamma-ray events and background events. For DM,
as an example we show in Fig.~\ref{fig:one} the case of a gravitino with mass $m_{3/2}=8$ GeV and lifetime 
$\tau_{3/2}=2.5 \times 10^{26}$ s.
Gamma-rays in the energy range between 3.4 GeV and 4.6 GeV are simulated.
This energy range corresponds to an interval of $\pm 2\ \Delta E$ 
around the position of the line (4 GeV), where $\Delta E$=0.3\ GeV is the 
energy resolution (at $68\%$ containment normal incidence) of the telescope at that energy.

Since we want to determine the Fermi-LAT capability of detecting local 
extragalactic DM structures, the gamma-ray emission from DM is what we
will refer to as {\it signal}.
For each direction of the sky $\psi$, the number of signal ($S$) and 
background ($B$) photons are determined integrating over a 3 $\times$ 3 degrees region centered in $\psi$. The $S/N$ ratio is then defined 
as $S/\sqrt{S+B}$.
%
%
From the map we can 
infer which extragalactic structure is the best target to derive our 
prospects for detecting gravitino DM.
Let us remark first that we are not simulating the contribution of the 
point sources already detected by Fermi-LAT and present in the 2 year catalog 
\cite{Collaboration:2011bm}. Since this contribution represents a source of 
uncertainty, the $S/N$ ratio should not be considered as a good estimator of 
the prospects for detecting DM in the regions where the contamination of the 
point sources is dominant. Also the discover in the
Fermi-LAT data of lobes structures \cite{Su:2010qj}, extending up to $50$ 
degrees above and below the galactic plane introduce a new source of 
background not included in this work. The inner galaxy region is characterized by a quite
large $S/N$ ratio (around 3). However, contamination from point sources as
well as the poorly-understood Fermi lobes is known to be large at low
galactic latitudes, significantly affecting the reliability of our predictions
in this region. It is worth noticing here that, as mentioned in the Introduction, in \cite{Abdo:2010nc, Bloom} the region $|b| \geq 10^\circ$ {\it plus} a $20^\circ \times 20^\circ$ square around the galactic center was used as target. Let us remark that these works follow a different kind of analysis, searching spectral deviations from a power-law behavior. We note that in such a large region of the sky the energy spectra used as background can be well described as a power law. Therefore, these works have no problems concerning background uncertainties.



Thus we neglect the zone with large $S/N$ close 
to the galactic center and plane, and focus only on the case of extragalactic 
sources. Among these, the object with the largest $S/N$ ratio ($S/N=2.1$) is 
the Virgo cluster. In our analysis we only select this cluster, indicated 
with a black square in Fig.\ref{fig:one}. 
Although in \cite{Cuesta:2010ex} filamentary regions
of the cosmic web were pointed out as good targets for DM decay searches, in this analysis, where
we are using a different particle physics model (and energy range), we do not find in principle a
significant $S/N$ ratio in those regions. 


As described in Ref.~\cite{Cuesta:2010ex}, the projected map is generated 
with Virgo being fixed in its real observed position because it is the best 
constrained object in the simulation, and therefore it is certainly the best 
object to consider for our purposes. Besides, there are two point sources 
detected near Virgo, M87 and 2E1228+1437, with an integrated flux of 
$(3.3 \pm 0.8)\times10^{-10}$ ph/cm$^2$s and 
$(1.6 \pm 0.6)\times10^{-10}$ ph/cm$^2$s between 3 and 10 GeV, respectively 
\cite{Collaboration:2011bm}. In a 5 $\times$ 5 degree region around
Virgo and free of the two point sources, the total gamma-ray flux from DM 
decay is $2.5 \times10^{-9}$ ph/cm$^2$s (for an example of a 8 GeV gravitino with a 
lifetime of $5 \times 10^{27}$s), one order of magnitude larger than the contribution of the 
point sources.
Smaller regions like $1\times 1$ or $3\times 3$ degrees are more affected by the emission 
due to the point sources. 
On the other hand, going to larger regions like $7\times 7$ degrees, the signal contribution is not significantly increasing and therefore the $S/N$ ratio decreases.
Thus, hereafter we select a $5\times 5$ degree region around Virgo as the best target to obtain our predictions.

\begin{figure}[t]
\includegraphics[width=0.7\textwidth]{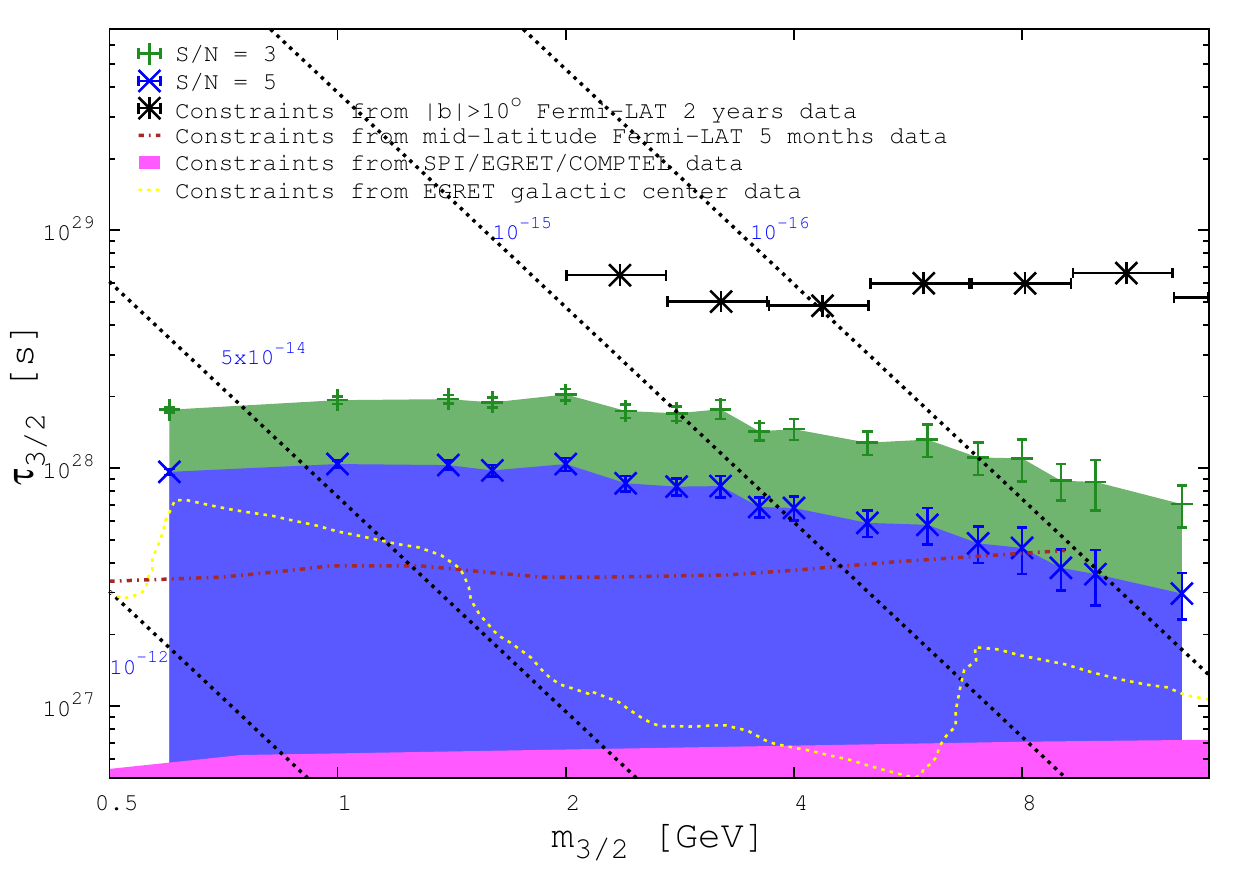}
\caption{\label{fig:two} Constraints on lifetime versus mass for gravitino DM in the $\mu\nu$SSM. Blue (green) points indicate values of $\tau_{3/2}$ and $m_{3/2}$ of the $\mu\nu$SSM gravitino corresponding to a detection of gamma-rays with a $S/N=5$ ($3$) in the $5 \times 5$ degree region centered on the position of the Virgo cluster, for a 5 years simulation using the Fermi Science Tools. The blue (green) region indicates points with $S/N$ larger than $5$ ($3$). The red dot-dashed line indicates the lower limit on $\tau_{3/2}$ obtained from the Fermi-LAT measurements of the mid-latitude gamma-ray diffuse emission after 5 months \cite{Choi:2009ng}. 
The yellow dashed line indicates the lower limit on  $\tau_{3/2}$ obtained from the EGRET measurements of the galactic center 
gamma-ray emission.
The black dots show the lower limit on $\tau_{3/2}$ obtained in the adopted energy bands \cite{Vertongen:2011mu}, from the Fermi-LAT measurements of the $|b| \geq 10^\circ$ gamma-ray diffuse emission after 2 years. 
The black dashed lines correspond to the predictions of the $\mu\nu$SSM \cite{Choi:2009ng} for several representative values $|U_{\tilde{\gamma}\nu}|^2$=$10^{-16}$, $10^{-15}$, $5\times 10^{-14}$, $10^{-12}$ (see Eq.~(\ref{master})). The magenta shaded region is excluded by gamma-ray observations such as SPI, COMPTEL and EGRET \cite{Yuksel:2007dr}.}
\end{figure}

With the purpose of scanning the most interesting portion of the $\mu\nu$SSM 
parameter space, we re-simulate the gamma-ray events from the region of 
5$\times$5 degrees around Virgo changing the value of the gravitino mass.
We run 17 different simulations of this region, each one with a different 
value for the gravitino mass, ranging from 0.6 to 10 GeV, for a given  
decay lifetime.
The lower bound of 0.6 GeV on the gravitino mass is chosen because it corresponds to a line energy of about 0.3 GeV,
where the PSF (point spread function) of the Fermi LAT becomes larger than our region of interest.
The energy interval covered by each simulation is 
$[(m_{3/2}/2-\Delta E),(m_{3/2}/2+\Delta E)]$, where the energy resolution (at $68\%$ containment normal incidence)
$\Delta E$ is computed at the position of the line.
Using the results of those simulations, we determine the values of lifetimes corresponding to a 
$S/N=5$ ($S/N=3$) .
These are plotted as blue (green) dots
in Fig. \ref{fig:two}, as a function of $m_{3/2}$. The blue (green) 
region indicates points with $S/N \geq 5$ ($3$). 
This is the main result of our work.

Let us remark that the errors in the figure are obtained propagating the 
statistical errors on the number of signal and background events (assuming 
Poissonian statistics). Regarding possible systematic errors, we note here 
that the points in Fig.~\ref{fig:two} are obtained with the 
P6\_V3\_DIFFUSE IRF but the simulations were repeated using both, 
P6\_V3\_DATACLEAN and P7CLEAN\_V6, and the results are found to be compatible. 
The use of P6\_V3\_DIFFUSE allows us to estimate the systematic error on our 
limits to be between 5 and 20\%.

As mentioned in the Introduction, in Ref.~\cite{Choi:2009ng} the area 
below the red dot-dashed line was disfavored by Fermi-LAT data of the
diffuse gamma-ray galactic emission in the mid-latitude range 
$10^\circ \le |b| \le 20^\circ$. 
In addition, in Ref.~\cite{Vertongen:2011mu}, the area below the black dots was also disfavored.
From a likelihood analysis focused on the region 
$|b| \geq 10^\circ$, lower bounds on $\tau_{3/2}$ of about $5 \times 10^{28}$ s 
were obtained in our region of interest below 10 GeV.
On the other hand, the area below the yellow dashed line is disfavored by the bounds obtained in \cite{kamion} by analyzing the data from EGRET in the galactic center region.
In particular, we used the upper limits on the gamma-ray line fluxes obtained in that work to constrain the $\mu\nu$SSM gravitino lifetime.
Finally, 
points in the magenta shaded region are excluded 
by gamma-ray observations from the galactic center obtained with the SPI spectrometer on INTEGRAL satellite, and the isotropic diffuse photon background as determined from SPI, COMPTEL and EGRET data \cite{Yuksel:2007dr}.

On the other hand,
the black dashed lines correspond to the predictions of the $\mu\nu$SSM 
for several representative values of the R-parity mixing parameter. As 
mentioned in Sect.~\ref{sec:two}, this is constrained to be 
$|U_{\widetilde{\gamma}\nu}|^{2}\sim 10^{-16}-10^{-12}$ in the $\mu\nu$SSM \cite{Choi:2009ng}, in order to reproduce the 
correct neutrino masses. 
As a consequence, any acceptable point must be in the area between the left and 
right black dashed lines.
Let us remark, however, that these bounds are very conservative, as discussed in \cite{Choi:2009ng}, and
in fact the results of a scan of the low-energy parameter space of the $\mu\nu$SSM implied that the range
$ 10^{-15}\leq |U_{\widetilde{\gamma}\nu}|^{2}\leq 5\times 10^{-14}$ 
is specially favored. The corresponding lines are also shown in the figure.

The combination of the constraints associated to red dot-dashed and black 
dashed lines, implies already that values of the gravitino mass larger than 
about 10 GeV are excluded, as well as lifetimes smaller than about 3 to 
5$\times 10^{27}$ s \cite{Choi:2009ng}. 
Actually, in the region of gravitino masses between 0.6 and about 1.5 GeV, 
lifetimes smaller than about 7 to 3 $\times 10^{27}$ s, respectively, are excluded because of the constraints associated to the
yellow dashed line. 
When constraints associated to black dots
are also imposed, it turns out that the gravitino mass has to be smaller than 
about 4 GeV,
and lifetimes have to be larger than about $6 \times 10^{28}$ s for gravitino masses between 2 and 4 GeV.
Thus, the combination of these results 
with the one obtained above for detection of DM from Virgo in 5 years of 
Fermi-LAT observations, leaves us with the blue and green areas above the 
yellow dashed and red dot-dashed lines, and gravitino mass smaller than 2 GeV, as those with good prospects for DM detection.
Summarizing, we find that a gravitino DM with a mass range of 
0.6--2 GeV, and with a lifetime 
range of about 
$3\times 10^{27}$--$2\times10^{28}$ s would be
detectable by the Fermi-LAT with a signal-to-noise ratio larger than 3.
If no gamma-ray lines are detected in 5 years, 
these regions of the parameter space of the $\mu\nu$SSM
would be excluded.

\section{Conclusions and outlook}
\label{sec:six}

In this work we have obtained the regions of the parameter space 
$(m_{3/2},\tau_{3/2})$ of the $\mu\nu$SSM with the best prospects for the 
detection of a gamma-ray monochromatic line from the decay of gravitino DM 
(see Fig.~\ref{fig:two}).
Summarizing, we find that a gravitino DM with a mass range of 
0.6--2 GeV, and with a lifetime 
range of about 
$3\times 10^{27}$--$2\times10^{28}$ s would be
detectable by the Fermi-LAT with a signal-to-noise ratio larger than 3.
We also obtain that gravitino masses larger than about 4 GeV are now disfavored in the $\mu\nu$SSM by Fermi-LAT data of the galactic halo.

In the analysis we have assumed 5 years of observation of the Virgo cluster 
by the Fermi-LAT space telescope. This cluster was 
selected as our optimal target due to its particularly high $S/N$ ratio. 
Of course, a more precise determination of the Fermi-LAT possibilities
of detecting gamma-ray lines towards Virgo would require the simulation
of both, M87 and 2E1228+1437. Also a more sophisticated analysis pipeline 
than the computation of the $S/N$ ratio, possibly through the determination 
of a test-statistics (TS) likelihood in order to derive the lower value of 
$\tau_{3/2}$ for which the signature of a line would be detectable with 
respect to the background.

Let us remark that the simulation of the gamma-ray flux was carried out with 
the use of the \texttt{gtobssim} routine from the Fermi Science Tools, 
whereas the DM distribution around the cluster has been modeled following 
the results of Ref.~\cite{Cuesta:2010ex} based on a constrained $N$-body 
simulation from the CLUES project \cite{CLUES}. With the present work we 
have also confirmed the potential of using extragalactic massive structures 
as optimal targets for decaying DM detection. For such a goal the maps of 
the local extragalactic DM distribution produced in Ref. \cite{Cuesta:2010ex} 
represent a unique, ready-to-use tool.


We conclude that there are good prospects for Fermi to detect monochromatic 
lines from gravitino decay in the energy range spanning from hundreds MeV 
to few GeV. That is also the energy range where it is more difficult to 
extract information from the data, due to imperfect parametrization of the 
background as a simple power law. Nevertheless, our results in 
Fig. \ref{fig:two} can be considered as an additional motivation to extend 
the Fermi-LAT analysis on lines to energies below 2 GeV.

\section*{Acknowledgments}
This work was supported by the Spanish MICINN's Consolider-Ingenio 2010 
Programme under grant MultiDark CSD2009-00064. The work of C. Mu\~noz and 
G.~A.~G\'{o}mez-Vargas was supported in part by MICINN under grants 
FPA2009-08958 and FPA2009-09017, by the Comunidad de Madrid under grant 
HEPHACOS S2009/ESP-1473, and by the European Union under the Marie 
Curie-ITN program PITN-GA-2009-237920. G. Yepes would like to thank the MICINN for financial support under
grants FPA 2009-08958, AYA 2009-13875-C03 and the
SyeC Consolider project CSD2007-00050. The simulations  used in this work were
performed at the Leibniz Rechenzentrum Munich (LRZ) and at
Barcelona Supercomputing Center (BSC).


\begin{thebibliography}{99}
\bibitem{Komatsu:2010fb}
  E.~Komatsu {\it et al.}  [WMAP Collaboration],
  Astrophys.\ J.\ Suppl.\  {\bf 192} (2011) 18
  [arXiv:1001.4538 [astro-ph.CO]].








\bibitem{reviews} For a review, see e.g.: C. Mu\~noz, Int. J. Mod. Phys. {\bf A19} (2004) 3093 [arXiv:hep-ph/0309346].



\bibitem{LopezFogliani:2005yw}
  D.~E.~L\'opez-Fogliani and C.~Mu\~noz,
  Phys.\ Rev.\ Lett.\  {\bf 97} (2006) 041801
  [arXiv:hep-ph/0508297].
  
\bibitem{reviewsmunu} For reviews, see:
  C.~Mu\~noz, 
  AIP Conf. Proc. {\bf 1200} (2010) 413 [arXiv:0909.5140 [hep-ph]]; 
  D. E.  L\'opez-Fogliani, arXiv:1004.0884 [hep-ph].
  
\bibitem{Escudero:2008jg}
  N.~Escudero, D.~E.~L\'opez-Fogliani, C.~Mu\~noz and R.~R.~de Austri,
  JHEP {\bf 12} (2008) 099
  [arXiv:0810.1507 [hep-ph]];
  P.~Ghosh and S.~Roy,
  JHEP {\bf 04} (2009) 069
  [arXiv:0812.0084 [hep-ph]];
  A.~Bartl, M.~Hirsch, A.~Vicente, S.~Liebler and W.~Porod,
  JHEP {\bf 05} (2009) 120
  [arXiv:0903.3596 [hep-ph]];
  J.~Fidalgo, D.~E.~L\'opez-Fogliani, C.~Mu\~noz and R.~Ruiz de Austri,
  JHEP {\bf 08} (2009) 105
  [arXiv:0904.3112 [hep-ph]];
   P.~Ghosh, P. Dey, B. Mukhopadhyaya and S.~Roy,
  JHEP {\bf 05} (2010) 087 [arXiv:1002.2705 [hep-ph]];
D.J.H. Chung and A. Long, Phys. Rev. {\bf D81} (2010) 123531
[arXiv:1004.0942[hep-ph]];
P. Bandyopadhyay, P.~Ghosh and S.~Roy,
arXiv:1012.5762[hep-ph]; S.~Liebler and W.~Porod, arXiv:1106.2921[hep-ph];
 J.~Fidalgo, D.~E.~L\'opez-Fogliani, C.~Mu\~noz and R.~Ruiz de Austri,
arXiv:1107.4614[hep-ph].

\bibitem{Choi:2009ng}
  K. Y. Choi, D. E. L\'opez-Fogliani, C. Mu\~noz and R. R. de Austri,
  JCAP {\bf 03} (2010) 028
  [arXiv:0906.3681 [hep-ph]].
  
  
  
  \bibitem{sneutrino} See e.g.: D.G. Cerde\~no, C. Mu\~noz and O. Seto,
Phys. Rev. {\bf D79} (2009) 023510 [arXiv:0807.3029 [hep-ph]], and references therein.


  
\bibitem{Atwood:2009ez}
  W. B. Atwood {\it et al.}  [Fermi LAT Collaboration],
  Astrophys.\ J.\  {\bf 697} (2009) 1071
  [arXiv:0902.1089];
  http://fgst.slac.stanford.edu/ .

\bibitem{Lapi:2009ee}
  A.~Lapi, A.~Paggi, A.~Cavaliere, A.~Lionetto, A.~Morselli and V.~Vitale,
  arXiv:0912.1766 [astro-ph.HE];
  V.~Vitale, A.~Morselli and f.~t.~F.~Collaboration,
  arXiv:0912.3828 [astro-ph.HE].
  
  \bibitem{navarro}
  J.F. Navarro, C.S. Frenk and S.D.M. White, 
  Astrophys. J. {\bf 462} (1996) 563. 
  

\bibitem{Abdo:2009mr}
  A.~A.~Abdo {\it et al.}  [Fermi LAT Collaboration],
  Phys.\ Rev.\ Lett.\  {\bf 103} (2009) 251101
  [arXiv:0912.0973 [astro-ph.HE]].
  
  
\bibitem{Choi:2010jt}
  K.~Y.~Choi, D.~Restrepo, C.~E.~Yaguna and O.~Zapata,
  JCAP {\bf 10} (2010) 033
  [arXiv:1007.1728 [hep-ph]].

 \bibitem{aurelio}
  M.A.~Diaz, S.~Garcia Saenz, B.~Koch,
  Phys. Rev. {\bf D84} (2011) 055007
  [arXiv:1106.0308 [hep-ph]].
  
\bibitem{Abdo:2010nc}
  A.~A.~Abdo {\it et al.} [Fermi LAT Collaboration],
  Phys.\ Rev.\ Lett.\  {\bf 104} (2010) 091302
  [arXiv:1001.4836 [astro-ph.HE]].

\bibitem{Bloom}
  E.~Bloom by the Fermi-LAT~Collaboration, talk at Aspen Winter Workshop 'Indirect and Direct Detection of Dark Matter', Feb. 6-8, 2011, http://www.slac.stanford.edu/exp/glast/aspen11 
  

  
  
\bibitem{Vertongen:2011mu}
  G.~Vertongen and C.~Weniger,
  JCAP {\bf 05 } (2011)  027.
  [arXiv:1101.2610 [hep-ph]].
  
\bibitem{kamion} 
A.R. Pullen, R.-R. Chary and M. Kamionkowski,
Phys. Rev. {\bf D76} (2007) 063006, Erratum-ibid. {\bf D83} (2011) 029904 [arXiv:1109.0512 [hep-ph]]



\bibitem{bilinear} 
D. Restrepo, M. Taoso, J.W.F. Valle and O. Zapata, arXiv:1109.0512 [hep-ph]



\bibitem{Markevitch:2001ri}
  M. Markevitch {\it et al.},
  Astrophys.\ J.\  {\bf 567} (2002) L27
  [arXiv:astro-ph/0110468].

\bibitem{Clowe:2006eq}
  D. Clowe, M. Bradac, A. H. Gonzalez, M. Markevitch, S. W.Randall, C. Jones and D. Zaritsky,
  Astrophys.\ J.\  {\bf 648} (2006) L109
  [arXiv:astro-ph/0608407].
  
  \bibitem{veryrecent}
 X. Huang, G.~Vertongen and C.~Weniger,
arXiv:1110.1529 [hep-ph].
 
  
\bibitem{Cuesta:2010ex}
  A.~J.~Cuesta {\it et al.}, Astrophys. J. {\bf 726} (2011) L6
  [arXiv:1007.3469 [astro-ph.HE]].
  
\bibitem{CLUES}
  http://www.clues-project.org 
  
  
\bibitem{Tools}  
  http://fermi.gsfc.nasa.gov/ssc/data/analysis
  
\bibitem{Takayama:2000uz}
  F.~Takayama and M.~Yamaguchi,
  Phys.\ Lett.\  B {\bf 485} (2000) 388
  [arXiv:hep-ph/0005214].
 
\bibitem{Buchmuller:2007ui}
  W.~Buchmuller, L.~Covi, K.~Hamaguchi, A.~Ibarra and T.~Yanagida,
  JHEP {\bf 03} (2007) 037
  [arXiv:hep-ph/0702184];
  G.~Bertone, W.~Buchmuller, L.~Covi and A.~Ibarra,
  JCAP {\bf 11} (2007) 003
  [arXiv:0709.2299 [astro-ph]];
  A.~Ibarra and D.~Tran,
  Phys.\ Rev.\ Lett.\  {\bf 100} (2008) 061301
  [arXiv:0709.4593 [astro-ph]];
  K.~Ishiwata, S.~Matsumoto and T.~Moroi,
  Phys.\ Rev.\ {\bf D78} (2008) 063505
  [arXiv:0805.1133 [hep-ph]].
   W.~Buchmuller, A. Ibarra, T. Shindou, F. Takayama and D. Tran,
  JCAP {\bf 09} (2009) 021
  [arXiv:0906.1187 [hep-ph]].

\bibitem{Prada:2004pi}
  F.~Prada, A.~Klypin, J.~Flix Molina, M.~Martinez, E.~Simonneau,
  Phys.\ Rev.\ Lett.\  {\bf 93 } (2004)  241301 [arxiv:astro-ph/0401512].
  
\bibitem{Overduin:2004sz}
  J.~M.~Overduin and P.~S.~Wesson,
  Phys. Rept.  {\bf 402} (2004) 267
  [arXiv:astro-ph/0407207].


\bibitem{Gottloeber:2010gv}
  S.~Gottloeber, Y.~Hoffman and G.~Yepes,
  arXiv:1005.2687 [astro-ph.CO].
  
  
\bibitem{Klypin} A.~Klypin, Y.~Hoffman, A.~Kravtsov and S.~Gottloeber,
  Astrophys.\ J.\  {\bf 596} (2003) 19
  [arXiv:astro-ph/0107104].
 
\bibitem{Cuesta_maps}
  http://www.clues-project.org/articles/darkmattermaps.html
  





\bibitem{Background}
  http://fermi.gsfc.nasa.gov/ssc/data/access/lat/BackgroundModels.html
  

\bibitem{Rando:2009yq}
  R.~Rando and f.~t.~F.~Collaboration,
  arXiv:0907.0626 [astro-ph.IM].
  

\bibitem{Collaboration:2011bm}
  T.~L.~Collaboration,
  arXiv:1108.1435 [astro-ph.HE].

\bibitem{Su:2010qj}
  M.~Su, T.~R.~Slatyer and D.~P.~Finkbeiner,
  Astrophys.\ J.\  {\bf 724} (2010) 1044
  [arXiv:1005.5480 [astro-ph.HE]].


 \bibitem{Yuksel:2007dr}
  H.~Yuksel and M.~D.~Kistler,
  Phys. Rev. {\bf D78} (2008) 023502
  [arXiv:0711.2906 [astro-ph]].
  










\end{thebibliography}
\end{document}